\documentclass[prl,twocolumn,aps,showpacs]{revtex4}
\usepackage{graphicx}
\begin{document}
\title{LDA+Gutzwiller Method for Correlated Electron Systems}

\author{XiaoYu Deng$^1$, Xi Dai$^1$, Zhong Fang$^2$}

\affiliation{$^1$Beijing National Laboratory for Condensed Matter
Physics and Institute of Physics, Chinese Academy of Sciences, Beijing
100080, China} 

\affiliation{$^2$International Center for Quantum Structure, Chinese
Academy of Sciences, Beijing, 100080, China}

\date{\today}   

\begin{abstract}
Combining the density functional theory (DFT) and the Gutzwiller
variational approach, a LDA+Gutzwiller method is developed to treat
the correlated electron systems from {\it ab-initio}. All variational
parameters are self-consistently determined from total energy
minimization. The method is computationally cheaper, yet the
quasi-particle spectrum is well described through kinetic energy
renormalization.  It can be applied equally to the systems from weakly
correlated metals to strongly correlated insulators. The calculated
results for SrVO$_3$, Fe, Ni and NiO, show dramatic improvement over
LDA and LDA+U.
\end{abstract}

 \maketitle

Despite of the successful stories of local-density-approximation (LDA)
for simple metals and band insulators, the applications of
first-principles calculations based on density functional theory (DFT)
are limited for correlated systems, due to insufficient treatment of
electron correlation. It has long been a challenge to have a simple
scheme which keeps the accuracy and parameter-free character of
DFT-type calculations, but can be equally applied to correlated and
non-correlated systems. Along this line, significant efforts have been
made in recent years to take into account explicitly the strong e-e
interaction~\cite{LDAU,LDASIC,DMFT}.  Among them, the LDA+U
method~\cite{LDAU} adopted the orbital dependent Hartree-like scheme,
and has been successfully applied to predict the insulating feature of
strongly correlated systems with long-range ordering. However, it
fails for intermediately correlated metallic systems.  The recently
developed LDA+DMFT method~\cite{DMFT} follow the same essence of LDA+U
scheme (i.e, the same effective Hamiltonian), however, the interaction
is treated by the proper calculation of frequency-dependent electron
self-energy through the dynamical mean field theory (DMFT). The method
is quite successful~\cite{DMFT}, but the frequency-dependency make the
method very expensive in practice. It is therefore important to
develop a new method which covers from the correlated metal to the
Mott insulator and at the same time is computationally as cheap as
LDA+U.

We will present in this paper that the Gutzwiller approach~\cite{Gut},
originally proposed to study the ferromagnetism in correlated band and
later widely used as analytical approach for correlated model
Hamiltonian~\cite{He3}, can be naturally combined with the DFT
formalism based on the variational principal. The idea is strongly
motivated by the recent progresses for the Gutzwiller approach: (1)
the original single band Gutzwiller method has been generalized to
multi-band case, and is exact for infinite lattice
dimensions~\cite{multiband-G}; (2) the Gutzwiller approach can
naturally cover both the noninteracting and the atomic
limit~\cite{limit-G}; (3) the post-LDA Gutzwiller treatment (starting
from the effective Hamiltonian extracted from LDA band) has been
successfully applied to some systems~\cite{Pu,Ni,NaCoO}. Having those
knowledge, we will show in the present paper that a fully
self-consistent LDA+Gutzwiller method can be developed, and it is easy
to be implemented in existing code and computationally
simple. Implementing the present scheme in our pseudo-potential
plane-wave code, we have calculated the non-magnetic metal SrVO$_3$,
magnetic metal Fe, Ni, and antiferromagnetic insulator NiO. The
results demonstrate significant improvement over LDA and LDA+U.

The basic idea of DFT and the Kohn-Sham equation is to find a
reference system (usually non-interacting system with single slater
determinant wave function $|\Psi_0\rangle$), which has the same charge
density but the kinetic energy can be explicitly computed. In such a
way, all unknown (both kinetic and potential) parts are moved into the
exchange-correlation energy $E_{xc}$. Unfortunately, the exact
functional form of $E_{xc}$ is not known, and the LDA for $E_{xc}$ is
not sufficient for the materials containing the partly filled narrow
bands, such as the $3d$ or $4f$ states. To improve this, the idea of
LDA+U or LDA+DMFT method is to correct the LDA Hamiltonian $H^{LDA}$
by including explicitly e-e interaction and remove the double counting
part from the LDA. Therefore in both LDA+U and LDA+DMFT, the
Hamiltonian reads,

\begin{equation}
H^{LDA+U}=H^{LDA}+\sum_{i\alpha \beta }U_{\alpha \beta }n_{i\alpha
}n_{i\beta }-H_{dc} 
\end{equation}

\noindent where $\left\vert i\alpha \right\rangle $ are a set of local
spin-orbitals with occupation number $n_{i\alpha}$ for lattice site
$i$, and $U_{\alpha\beta}$ gives the interaction strength between the
orbitals, $H_{dc}$ is the double couting term.

Now the problem to be solved is to minimize the total energy through
variational principal. In LDA+U, this is done with respect to the
single Slatter determinent trial wave function $|\Psi_0\rangle$, which
can only improve the interaction part of the total energy and leave
the kinetic part unchanged. In LDA+DMFT, however, both the single
particle wave functions and the spectral density of the local orbitals
are considered to be the variational parameters. The enlargement of
the variational space makes the LDA+DMFT work quite well for both
interaction part and kinetic part of the total energy. But the price
of LDA+DMFT is that we have to deal with the frequency dependence of
the self energy for each local orbitals, which makes it very
expensive.

An alternative way for the problem is to improve the trial
wave-function such that the kinetic part can be treated as better as
possible. This is what we will done in the following. It is known that
$|\Psi_{0}\rangle$ is not good, and a better candidate for the trial
wave-function is the Gutzwiller wave-function~\cite{Gut}, which is
basically the lattice version of the Jastrow type wave function for
quantum liquid and has been extensively used in the strongly
correlated systems~\cite{He3}. It is defined as

\begin{equation}
|\Psi^{G}>=\hat{P}|\Psi^0>
\end{equation}

The projection operator $\hat{P}$ is used to reduce the weight of
configurations for the local orbitals with relatively high energy, and
it reads

\begin{equation}
\hat{P}=\prod\limits_{i\Gamma }\left[ 1+\left( \lambda _{i\Gamma }-1\right)
\left\vert i,\Gamma \right\rangle \left\langle i,\Gamma \right\vert \right] 
\end{equation}

\noindent where $\Gamma $ denotes the $\Gamma $-th many-body
configuration for the local orbitals in given unit cell $i$ and
$\lambda _{i\Gamma }$ are the variational parameters describing the
weight for given configuration $\Gamma $. Therefore our task is to
minimize the total energy with respect to the Gutzwiller
wave-function:

\begin{equation}
E^{LDA+G}(\rho )=\langle\Psi _{G}|H^{LDA}|\Psi
_{G}\rangle-E_{dc}+\langle\Psi _{G}|H_{U}|\Psi _{G}\rangle
\label{expect}
\end{equation}

Thanks to the recent progresses for Gutzwiller approach, although the
exact solution of the above expectation is unknown, the Gutzwiller
approximation (GA) is a very tractable approximation for the multiband
Hubbard model and has been proven to be exact for the infinite
dimension. In the present paper, we apply the GA directly to the LDA
Hamiltonian rather than tight binding form extracted from it. To do
so, we only need a set of local orbitals $|i\alpha\rangle$ where we
implement the local correlation effect. Following reference~\cite{Pu},
we can derive the total energy under GA as,

\begin{equation}
E^{LDA+G}(\rho )\approx <\Psi _{0}|H_{eff}^{LDA+G}|\Psi
_{0}>-E_{dc}+\sum_{i,\Gamma }E_{\Gamma }^{loc}m_{\Gamma }^{2} 
\end{equation}

\noindent with the Gutzwiller effective Hamtonian in momentum ($k$)
space given as

\begin{eqnarray}
H_{eff}^{LDA+G}&=&[\hat{Q}+(1-\hat{R})]H^{LDA}[\hat{Q}+(1-\hat{R})]  \nonumber \\
&&+\sum_{\alpha k}(1-q_{\alpha }^{2})\epsilon _{\alpha }^{0}\left\vert
\alpha k\right\rangle \left\langle \alpha k\right\vert 
\end{eqnarray}

\begin{equation}
\hat{Q} =\sum_{\alpha k}q_{\alpha }\left\vert \alpha k\right\rangle
\left\langle \alpha k\right\vert, \ \ \ 
\hat{R} =\sum_{\alpha k}\left\vert \alpha k\right\rangle \left\langle
\alpha k\right\vert
\end{equation}

\noindent where $E_{\Gamma }$ is the eigen value and $m_{\Gamma
}^{2}$\ is the weight of the $\Gamma$-th configuration,
$\epsilon_\alpha^0$ is the on-site energy of each local orbital. The
key quantities in our formalism are $q_{\alpha }$ ($0<q_\alpha<1$),
which describe the kinetic energy renormalization of local orbitals,
and can be expressed in terms of $m_{\Gamma }$ through

\begin{equation}
q_{\alpha }=\sum_{\Gamma \Gamma ^{\prime }}\left\langle \Gamma
^{\prime }\right\vert C_{\alpha }^{\dag }\left\vert \Gamma
\right\rangle \frac{ m_{\Gamma }m_{\Gamma ^{\prime }}}{\sqrt{n_{\alpha
}\left( 1-n_{\alpha }\right) }}
\end{equation}

The main strategy of above formalism is that, kinetic energy of
correlated orbitals are renormalized by factor $q_\alpha$, while all
other non-correlated orbitals are taken into account by $(1-\hat{R})$
in the complete local basis projection.  Therefore under GA, the total
energy can be expressed again as the functional of the non-interacting
wave function $\Psi_{0}$, but with additional variational parameters
of the weighting factors ${ m_{\Gamma }}$, along with the following
necessary constraint

\begin{equation}
\sum_{\Gamma }\left\langle \Gamma \right\vert C_{\alpha }^{\dag }C_{\alpha
}\left\vert \Gamma \right\rangle m_{\Gamma }^{2}=n_{\alpha
}=\sum_{k}\left\langle \Psi _{0}\right\vert C_{\alpha k}^{\dag }C_{\alpha
k}\left\vert \Psi _{0}\right\rangle 
\end{equation}

\noindent the original Gutzwiller variational parameters $\lambda
_{i\Gamma }$ can be obtained by $\lambda _{\Gamma }=\frac{m_{\Gamma
}}{m_{\Gamma }^{0}}$, with $ m_{\Gamma }^{0}=\left\langle \Psi
_{0}\right\vert \left\vert \Gamma \right\rangle $.

Having above equations, it is easy to obtain Kohn-Sham like equations
by the variational method,

\begin{equation}
\frac{\partial{E^{LDA+G}(\rho)}}{\partial{\Psi_0}}=0, \ \ \ 
\frac{\partial{E^{LDA+G}(\rho)}}{\partial{m_\Gamma}}=0
\end{equation}

The charge density under GA can be constructed as:

\begin{eqnarray}
\rho^{LDA+G}(r)&=&\left\langle \Psi _{0}\right\vert
[\hat{Q}+(1-\hat{R})] | r \rangle \langle r |
[\hat{Q}+(1-\hat{R})]\left\vert \Psi _{0}\right\rangle \nonumber \\
&+&\sum_{\alpha k}(1-q_{\alpha }^{2})\left\langle \Psi _{0}|\alpha
k\right\rangle \left\langle \alpha k|\Psi _{0}\right\rangle
\rho_{\alpha k}(r)
\end{eqnarray}

In practice, two steps are followed. (1) For fixed $m_\Gamma$, the
$|\Psi_0\rangle$ is optimized by solving the effective Kohn-Sham
equations. This step is basically the same as all other LDA
calculations, with additional little cost for the projection to local
orbitals; (2) For fixed $|\Psi_{0}\rangle$, $m_{\Gamma }$ are
optimized by solving set of linear equations~\cite{SCF-GUT}, which
have dimension of number of configurations $\Gamma$ (for instance
$2^{10}$ for five $d$ orbitals, and the number of configurations can
be reduced significantly by considering the symmetry). The second step
is addition to the LDA or LDA+U scheme, however, the computational
cost is rather small compared with the cost by wave function
optimization (i.e., the first step), because only set of linear
equations needed to be solved~\cite{SCF-GUT}. The final solution of
the system is obtained until the charge density (and total energy) is
self-consistently converged.

The above proposed LDA+Gutzwiller method was implemented in our BSTATE
(Beijing Simulational Tool for Atom Technology)
code~\cite{STATE-review}, which uses plane-wave ultra-soft
pseudo-potential method. For the local basis, the physical choice is
the wannier function constructed from the atomic orbital. This was
done by using the projected wannier function method~\cite{PWANF} in
our calculations. For the interaction term, the on-site
density-density interactions are considered in the present
studies. Several typical systems have been calculated, and the results
are discussed below.

\begin{figure}
\includegraphics[scale=0.43]{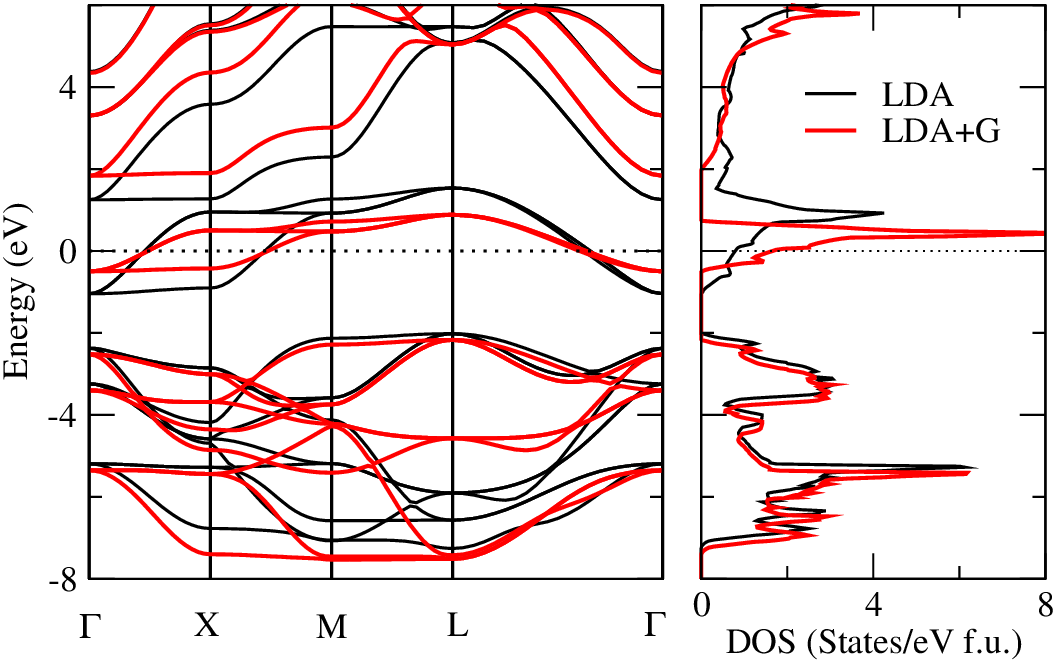}
\caption{The calculated band structure and density of states for
SrVO$_3$ using LDA and LDA+Gutzwiller method.}
\end{figure}

\noindent{\it 1. Non-magnetic correlated metal: SrVO$_3$}

SrVO$_3$ is a correlated metal with $3d$-$t_{2g}^1$ configuration. The
simple cubic perovskite crystal structure and non-magnetic electronic
state make it an idea test material~\cite{SVO-theory}. The
one-particle spectrum calculated with LDA is poorly compared with
experiments~\cite{SVO-exp}. The main problems are: (1) the calculated
band width is about 40\% wider than photoemission observation; (2) the
estimated effective mass is about 2-3 times lower than experimental
results from specific heat and susceptibility; (3) the photoemission
peak observed around 2eV below Fermi level ($E_F$) is not understood.

However, most of the features can be understood from our
LDA+Gutzwiller calculations. We choose the $3d$ Wannier function as
the interacting local orbits with effective interaction energy
$U$=5.0eV following the literatures~\cite{SVO-theory}. Fig.1 shows the
calculated band structure and the density of states of quasi-particle
spectrum. It is clear that the band width is reduced by about 40\%
compared with LDA. From the calculated quasi-particle spectrum it is
straight forward to calculate the effective mass enhancement
($m^*/m$), which is 2.1 times larger than LDA. All the results are in
good consistency with experiments~\cite{SVO-exp}, and much better than
that obtained from LDA. Finally, it is worth to make the comment:
since all the quasi particle part has been well treated in the present
scheme (incoherent part is not included yet), and no DOS is found
around -2eV region, it further suggests that the observed
photoemission peak in this region should be incoherent as suggested by
other studies~\cite{SVO-theory}.

\begin{figure}
\includegraphics[scale=0.45]{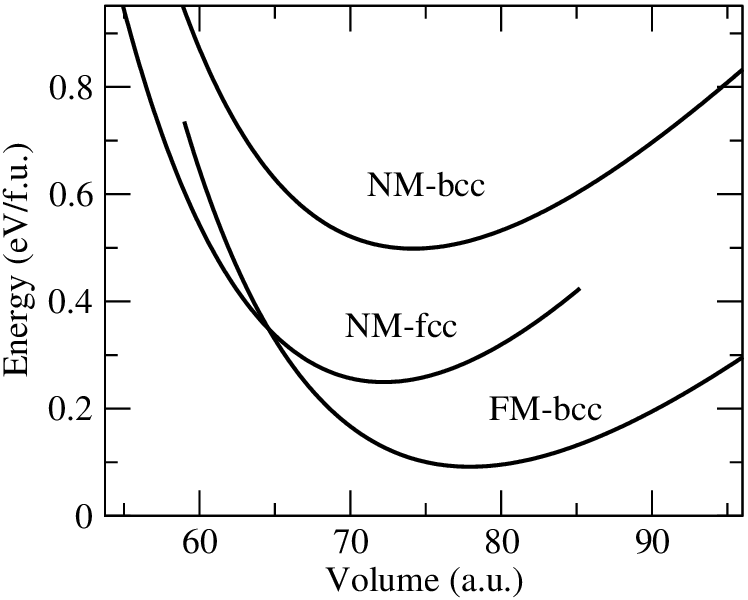}
\caption{The calculated total energy as function of volume for
different phases of bulk Fe by LDA+Gutzwiller method. The bcc FM
ground state is correctly predicted.}
\end{figure}

\noindent{\it 2. Magnetic correlated metal: Fe and Ni}

Bulk Fe and Ni are typical magnetic metal with intermediate
correlations. For Fe, the LDA fails to predict the bcc FM ground
state, although GGA correctly do so. Both LDA and GGA overestimate the
band width by about 10-20\% compared with
experiements~\cite{Fe-AREPS}. For the Ni, the problems are more
serious, the band width is overestimated by about 30\%, and the spin
polarization is hardly compared with experiments~\cite{Ni}. For such
intermediately correlated metal, the LDA+U method may improve one or
two discrepancies to some extent, while the price to be payed is that
other predicted properties become even worse than LDA.

For our calculations, we first determine the effective $U$ from
constrained-LDA calculations to be 7.0eV (Fe) and 9.0eV (Ni) for the
choice of our local orbital, and Hund's coupling $J$ is fixed to be
1.0eV (a common choice for $3d$ elemental metal). The results,
summarized in Table I and Fig.2, show that most of the discrepancies
are systematically corrected compared with experiments, suggesting the
advantages of present scheme. In particular, the following
improvements are significant: (1) the band width renormalization is
correctly predicted and effective mass enhancement (specific heat
coefficient) can be well compared with experiments. (2) the calculated
lattice parameter, bulk modulus and magnetic moment show systematic
improvement. Finally, we want to point out that for the down-spin band
of Ni, around the X point near $E_F$, one of the bands predicted by
LDA to be above $E_F$, is now correctly predicted to be below $E_F$,
in consistency with recent AREPS data~\cite{Ni}.

\begin{figure}
\includegraphics[scale=0.42]{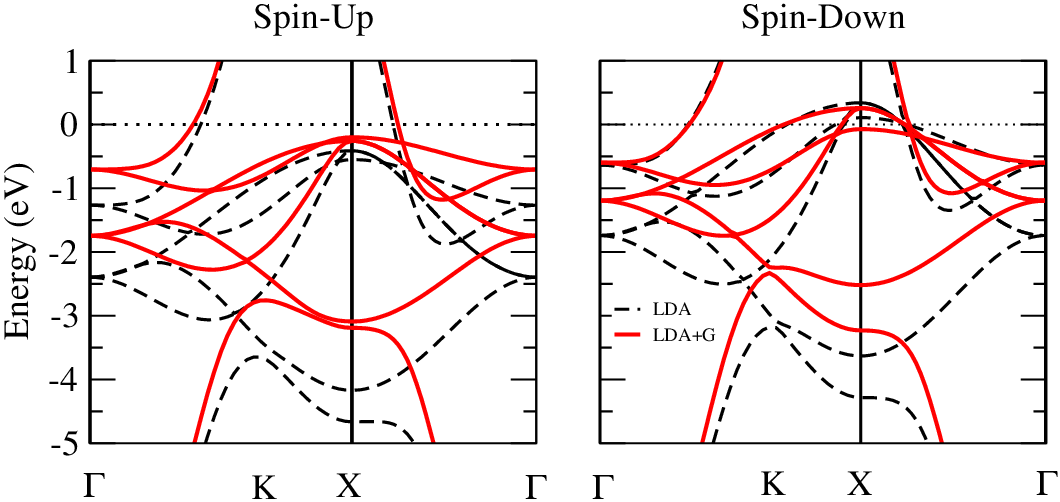}
\caption{The calculated band structure of fcc FM Ni.}
\end{figure}

\begin{table}
\caption{The calculated property parameters for bcc FM Fe and fcc FM
Ni in comparison with experimental results. They are equilibrium
lattice constant $a_0$, bulk modulus $B$, spin magnetic moment $M$,
specific heat coefficient $\gamma$, and the occupied energy band width
$W$. The experimental data are from Ref.~\cite{Fe-Ni-Exp}.}
\begin{tabular}{c|l|c|c|c|c|c}
\hline
     &      &a$_0$(bohrs)  &$B$(GPa)   &$M$($\mu_B$) &$\gamma$($\frac{mJ}{k^2mol}$) &$W$(eV) \\ \hline
     &LDA   &5.21          &227      &2.08       &2.25           &3.6    \\
Fe   &LDA+G &5.39          &160      &2.30       &3.52           &3.2    \\ 
     &Exp.  &5.42          &168      &2.22       &3.1,3.69       &3.3     \\ 
\hline
     &LDA   &6.49          &250      &0.59       &4.53           &4.5     \\
Ni   &LDA+G &6.61          &188      &0.50       &6.9            &3.2     \\
     &Exp.  &6.65          &186      &0.42,0.61  &7.02           &3.2    \\  \hline
\end{tabular}
\end{table}

\noindent{\it 3. AF correlated Insulator: NiO}

In the present LDA+Gutzwiller scheme, both effective on-site level
renormalization and kinetic renormalization are included, while only
the former is considered in LDA+U scheme. For the intermediately
correlated metallic systems, the kinetic energy renormalization is
significant, this is reason why LDA+U scheme fails. However, for the
large U limit, where effective on-site level renormalization dominate,
such as the AF correlated insulator NiO, our calculations suggest that
the kinetic energy renormalization factor $q_\alpha$ is almost
utility, and the obtained electronic structure from LDA+Gutzwiller
method is almost the same as that of LDA+U. However, it should be
noted that this is true only for the cases where $q_\alpha$ close to
1. For some of the AF insulators, the localized moment is far away
from utility, we expect that the kinetic energy renormalization will
also contribute. In these cases, the present LDA+Gutzwiller approach
will give a better results than LDA+U even for the AF ordered
insulators. We will leave this issue for future studies.

In summary, we have shown that the Gutzwiller approach can be well
combined with the DFT through variational principal. As the results, a
fully self-consistent LDA+Gutzwiller method is developed. The method
not only keeps the accuracy and parameter-free character of LDA-type
calculations (such as the total energy calculations), but also
automatically cover the region from weakly to strongly correlated
systems. The quasi particle spectrum is properly described by taken
into account the kinetic energy renormalization, and the calculated
results for several typical systems demonstrate the strong advantage
over LDA and LDA+U results. On the other hand, the scheme itself is
easy to be implemented and computationally much cheaper than the
LDA+DMFT method.

We acknowledge the supports from NSF of China (No.10334090, 10425418,
60576058), and that from the 973 program of China (No.2007CB925000).

\end{document}